\documentclass[12pt]{article}
\pdfoutput=1 

\usepackage{graphicx}

\usepackage{pbox}
\usepackage{epsfig}
\usepackage{rotating}
\usepackage{color, colortbl}
\usepackage[normalem]{ulem}
\usepackage{multirow}
\usepackage{tabularx}
\usepackage{booktabs}
\usepackage{graphicx}
\usepackage{lipsum}
\usepackage{dcolumn}
\usepackage{hyperref}
\usepackage{float}
\usepackage{amsmath}
\usepackage{amssymb}
\usepackage{ulem}
\usepackage{color}

\newcommand{\red}[1]{\color{red} #1 \color{black}}

\newcommand{\AddrAHEP}{
  {\it AHEP Group, Instituto de F\'{\i}sica Corpuscular --
    C.S.I.C./Universitat de Val{\`e}ncia \\
    Edificio de Institutos de Paterna,
 C/Catedratico Jos\'e Beltr\'an, 2 E-46980 Paterna (Val\`{e}ncia) - Spain}}

\begin{document}

\begin{titlepage}

\begin{center}

{\bf Sensitivities to neutrino electromagnetic properties\\ at the TEXONO experiment}

\vspace*{.8cm}

{T.S. Kosmas~$^1$}\footnote{hkosmas@uoi.gr}, 
{O.G. Miranda~$^2$}\footnote{omr@fis.cinvestav.mx}, 
{D.K. Papoulias~$^{1,3}$}\footnote{dimpap@cc.uoi.gr}, \\
{M. T\'ortola~$^3$}\footnote{mariam@ific.uv.es}, and 
{J.W.F. Valle~$^3$}\footnote{valle@ific.uv.es, URL:
  http://astroparticles.es/} \\[.3cm]

 {\it $^1$~Division of Theoretical Physics, 
 University of Ioannina,\\GR-45110 Ioannina, Greece} \\
{\it$^2$~Departamento de F\'{\i}sica, Centro de Investigaci\'on
  y de Estudios Avanzados del IPN,\\ Apdo. Postal 14-740 07000 Mexico,
  DF, Mexico} \\
{\it$^3$~\AddrAHEP} 

\vspace*{.8cm}

\begin{abstract}
  
  The possibility of measuring neutral-current coherent elastic
  neutrino-nucleus scattering (CENNS) at the TEXONO experiment has
  opened high expectations towards probing exotic neutrino
  properties. Focusing on low threshold Germanium-based targets with
  kg-scale mass, we find a remarkable efficiency not only for
  detecting CENNS events due to the weak interaction, but also for
  probing novel electromagnetic neutrino interactions. Specifically,
  we demonstrate that such experiments are complementary in performing
  precision Standard Model tests as well as in shedding light on
  sub-leading effects due to neutrino magnetic moment and neutrino
  charge radius. This work employs realistic nuclear
  structure calculations based on the quasi-particle random phase
  approximation (QRPA) and takes into consideration the crucial 
  quenching effect corrections. Such a treatment, in 
  conjunction with a simple statistical
  analysis, shows that the attainable sensitivities are improved by one
  order of magnitude as compared to previous studies.

\end{abstract}
\end{center}

\smallskip
\noindent \textit{Keywords:} Reactor neutrinos, Coherent elastic neutrino-nucleus scattering   (CENNS), Weak mixing angle, Neutrino magnetic moment, Neutrino charge radius, Quenching factor

\end{titlepage}

\section{Introduction}
\label{sec:intro}

The robust discovery of neutrino oscillations in the propagation of
solar and atmospheric neutrinos~\cite{Maltoni:2004ei,Nunokawa:2007qh},
confirmed at accelerator and reactor neutrino
sources~\cite{Tortola:2012te,Forero:2014bxa} has provided us with a
rather solid proof for the existence of neutrino masses and
mixing~\cite{Schechter:1980gr,Schechter:1981cv} and hence the clearest
evidence for the need of physics beyond the Standard Model (SM)~\cite{Kosm-talk-Finland}.
These results have prompted a great rush to produce adequate SM
extensions with small neutrino masses~\cite{Boucenna:2014zba}.

Underpinning the ultimate origin of neutrino mass stands out as one of
the biggest challenges in particle physics~\cite{Valle:2015pba}.
A generic feature of many such schemes is the presence of
non-vanishing neutrino electromagnetic (EM)
properties~\cite{Schechter:1981hw,Shrock:1982sc,kayser:1982br,Nieves:1981zt,Beacom:1999wx,Broggini:2012df}.
While the neutrino masses indicated by oscillation data are perhaps
too small to induce sizeable magnetic moments, this issue is rather
model dependent, and one cannot exclude this possibility on general
grounds~\cite{Papoulias:2015iga}.
If large enough, these may still play an important sub-leading role in
precision neutrino studies~\cite{Beringer:1900zz}, despite the good
agreement found within the three-neutrino oscillation picture.
Non-zero diagonal magnetic moments exist for massive Dirac neutrinos.
In contrast, in the general Majorana neutrino case all magnetic moments are
transition-type. Therefore, the study of neutrino magnetic moments
would be a powerful tool towards distinguishing their Dirac or
Majorana character~\cite{Grimus:2002vb,Tortola:2004vh}.

The detection of neutral-current (NC) coherent elastic neutrino-nucleus
scattering (CENNS) processes by measuring the nuclear recoil spectrum
of the scattered nucleus has by now become
feasible~\cite{Scholberg:2005qs}. As a concrete example, the newly
formed COHERENT Collaboration at the Spallation Neutron Source (SNS)
has excellent prospects~\cite{Bolozdynya:2012xv,Akimov:2013yow},
motivating also theoretical
effort~\cite{Papoulias:2015vxa,Kosmas:2015sqa}.
In this work, we consider the possibility of revealing signs of new
physics through a detailed study of CENNS at the TEXONO
experiment~\cite{Wong:2006nx,Deniz:2009mu,Soma:2014zgm}. We
demonstrate that the use of sub-keV Germanium-based kg-scale
detectors~\cite{Wong:2005vg,Wong:2010zzc,Chen:2014ypv}, provides a favourable
experimental set up with good prospects for performing precision
SM tests, as well as probing EM neutrino
properties~\cite{Vogel:1989iv}, such as the neutrino magnetic
moment~\cite{Grimus:2002vb,Tortola:2004vh,Miranda:2003yh,miranda:2004nz}
and the neutrino charge
radius~\cite{Bardeen:1972vi,Lee:1973fw,Barranco:2007ea,Hirsch:2002uv,bernabeu:2000hf}.
The total number of events expected in an experiment searching for
CENNS depends strongly on the energy threshold $T_{thres}$, as well as
the total mass of the
detector~\cite{Papoulias:2015vxa,Kosmas:2015sqa}. For low energy
thresholds and more massive detectors, the total number of events
expected is significantly larger and, therefore, the attainable
sensitivities are higher. In addition, this work highlights that the present calculations become more realistic by considering  quenching effect corrections~\cite{Simon:2002cw,Giomataris:2005fx,Vergados:2009ei}. The sensitivity is evaluated by assuming
that a given experiment searching for CENNS events will measure
exactly the SM expectation. Thus, any
deviation~\cite{Miranda:2004nb,Barranco:2005yy, Barranco:2005ps} is
understood as a signature of new
physics~\cite{Barranco:2007tz,Escrihuela:2009up,Escrihuela:2011cf,Papoulias:2013gha,Miranda:2015dra}.

Apart from the possibility of the first ever detection of CENNS
events, our present results emphasise the potentiality of discovering
neutrino interactions beyond the SM expectations~\cite{Giunti:2015gga}.  In our
estimates we perform nuclear structure calculations within
the context of the quasi-particle random phase approximation (QRPA) that uses realistic nuclear forces~
\cite{Kosmas:1994ti,Kosmas:2001ia,Chasioti2009234,Tsakstara:2011zzd,Giannaka:2015sta},
and employ a $\chi^2$-type statistical analysis. We find that the
prospects for improving current bounds on $\mu_{\nu_e}$ are rather
promising and complementary to future sensitivities on the muon
neutrino magnetic moment, $\mu_{\nu_\mu}$~\cite{Kosmas:2015sqa}.

\section{Coherent elastic neutrino-nucleus scattering}

The coherent elastic scattering of neutrinos upon a nucleus is
described within the SM starting from the neutrino-quark
NC interaction Lagrangian. However, as mentioned
above, one has good reasons to expect corrections coming from new
physics~\cite{Maltoni:2004ei}, such as non-standard
interactions (NSI)~\cite{Miranda:2004nb,Barranco:2005yy,
  Barranco:2005ps,Barranco:2007tz,
  Escrihuela:2009up,Escrihuela:2011cf,Miranda:2015dra} or non-trivial
neutrino EM properties
~\cite{Schechter:1981hw,Shrock:1982sc,kayser:1982br,Nieves:1981zt,Beacom:1999wx,Broggini:2012df,Papoulias:2015iga,Miranda:2003yh,miranda:2004nz,Barranco:2007ea}.

\subsection{Standard model prediction}
\label{sec:stand-model-pred}

Within the context of SM, for low energies ($E_\nu \ll M_W$)
accessible to neutrino experiments, the weak neutral-current CENNS can
be naturally studied by considering the $V \pm A$ interaction of
four-fermion $\nu \nu f f$ type operators entering the effective
Lagrangian, $\mathcal{L}_{\mathrm{SM}}$, written as~\cite{Barranco:2005yy}
\begin{equation}
\mathcal{L}_{\mathrm{SM}} = - 2 \sqrt{2} G_{F} \sum_{ \begin{subarray}{c}  P = L,R \\ f= \, u,d \\ \alpha= e, \mu,\tau 
\end{subarray}} g_{\alpha \alpha}^{f,P}   
\left[ \bar{\nu}_{\alpha} \gamma_{\rho} L \nu_{\alpha} \right] \left[ \bar{f} \gamma^{\rho} P f \right] \, .
\label{SM_Lagr}
\end{equation}
In this Lagrangian, the chiral structure of the SM weak interaction is
expressed by the projectors $P=\{L,R\}$, while $\alpha$ denotes the
neutrino flavour and $f$ refers to the first-generation quarks.  The
relative strength of the left- and right-handed couplings for $u$- and
$d$-quark to the $Z$-boson with respect to the Fermi constant $G_F$
are given as
\begin{equation}
\begin{aligned}
g_{\alpha \alpha}^{u,L} =& \rho_{\nu N}^{NC} \left( \frac{1}{2}-\frac{2}{3} \hat{\kappa}_{\nu N} \hat{s}^2_Z \right) + \lambda^{u,L} \, ,\\
g_{\alpha \alpha}^{d,L} =& \rho_{\nu N}^{NC} \left( -\frac{1}{2}+\frac{1}{3} \hat{\kappa}_{\nu N} \hat{s}^2_Z \right) + \lambda^{d,L} \, ,\\
g_{\alpha \alpha}^{u,R} =& \rho_{\nu N}^{NC} \left(-\frac{2}{3} \hat{\kappa}_{\nu N} \hat{s}^2_Z \right) + \lambda^{u,R} \, ,\\
g_{\alpha \alpha}^{d,R} =& \rho_{\nu N}^{NC} \left(\frac{1}{3} \hat{\kappa}_{\nu N} \hat{s}^2_Z \right) + \lambda^{d,R} \, .
\end{aligned}
\end{equation}
In the latter expressions, after including the relevant radiative
corrections, we have $\hat{s}^2_Z = \sin^2 \theta_W= 0.23120$,
$\rho_{\nu N}^{NC} = 1.0086$, $\hat{\kappa}_{\nu N} = 0.9978$,
$\lambda^{u,L} = -0.0031$, $\lambda^{d,L} = -0.0025$ and
$\lambda^{d,R} =2\lambda^{u,R} = 7.5 \times
10^{-5}$~\cite{Beringer:1900zz}.  From the effective Lagrangian in
Eq.~(\ref{SM_Lagr}), the differential cross section with respect to
the nuclear recoil-energy, $T$, for the case of a CENNS off a
spherical spin-zero nucleus of mass $M$,
reads
%

\begin{equation}
\left( \frac{d\sigma}{dT} \right)_{\mathrm{SM}} = \frac{G_F^2 \,M}{2 \pi} \left[1- \frac{M T}{E_\nu^2} + \left(1- \frac{T}{E_\nu} \right)^2
\right] \left\vert\langle gs \vert\vert \hat{\mathcal{M}}_0(q) \vert\vert gs \rangle\right \vert^2\, .
\label{SM_dT}
\end{equation}

For $gs\rightarrow gs$ transitions, the corresponding coherent nuclear
matrix element takes the form~\cite{Papoulias:2013gha,Papoulias:2015vxa}
\begin{equation}
\begin{aligned}
\langle gs \vert\vert \hat{\mathcal{M}}_0(q) \vert\vert gs \rangle = &
 \left[ 2(g_{\alpha \alpha}^{u,L} + g_{\alpha \alpha}^{u,R}) + (g_{\alpha \alpha}^{d,L} + g_{\alpha \alpha}^{d,R}) \right] Z F_Z (q^2) \\
 + & \left[ (g_{\alpha \alpha}^{u,L} +
g_{\alpha \alpha}^{u,R}) +2(g_{\alpha \alpha}^{d,L} + g_{\alpha
  \alpha}^{d,R}) \right] N F_N (q^2)  \, .
  \end{aligned}
\label{SM-ME}
\end{equation} 
Note that, due to the smallness of the coupling of protons with the
Z-boson, the main contribution to the CENNS cross section essentially
scales with the square of the neutron number $N$ of the target nucleus
(see e.g.~\cite{Papoulias:2013gha}).  We stress that the differential
cross section is evaluated with high significance by weighting the
nuclear matrix element with corrections provided by the proton
(neutron) nuclear form factors $F_{Z(N)}(q^2)$. This way the finite
nuclear size is taken into account with respect to the typical
momentum transfer, $q \simeq \sqrt{2 M T}$. Furthermore, the $N^2$
enhancement of the CENNS cross section makes the relevant experiments
favourable facilities to probe the neutron form factor of the target
nucleus at low
energies~\cite{Scholberg:2005qs,Bolozdynya:2012xv,Akimov:2013yow}.

From a nuclear theory point of view, the reliability of the present
CENNS cross sections calculations is maximised in terms of accuracy by
performing nuclear structure calculations in the context of QRPA~\cite{Kosmas:2001ia,Chasioti2009234}.  Motivated by
its successful application on similar calculations for various
semi-leptonic nuclear processes~\cite{Tsakstara:2011zzd,Giannaka:2015sta},
in this work we construct explicitly the nuclear ground state, $\vert
gs\rangle \equiv \vert 0^+\rangle$, of the relevant even-even isotope
through the solution of the BCS equations (for a detailed description
see Ref.~\cite{Papoulias:2015vxa}).

\subsection{Electromagnetic neutrino-nucleus cross sections}

The existence of neutrino masses is well-established thanks to the
current neutrino oscillation data, implying that they could have
exotic properties, such as non-zero neutrino magnetic moments. In this
framework, potential neutrino-nucleus interactions of EM nature have
been
considered~\cite{Schechter:1981hw,Shrock:1982sc,kayser:1982br,Nieves:1981zt,Beacom:1999wx,Broggini:2012df},
resulting in corrections to the weak CENNS cross section of the
form~\cite{Vogel:1989iv}
\begin{equation}
\left( \frac{d \sigma}{dT}\right)_{\mathrm{tot}} = \left( \frac{d \sigma}{dT} \right)_{\mathrm{SM}} 
+ \left( \frac{d \sigma}{dT} 
\right)_{\mathrm{EM}}\, .
\label{tot-crossec}
\end{equation}
Here, the helicity-violating EM contribution to the neutrino-nucleus
cross section can be parametrised in terms of the proton nuclear form
factor, the fine structure constant $a_{em}$ and the electron mass $m_e$
as~\cite{Papoulias:2015iga}
\begin{equation}
\left( \frac{d \sigma}{dT} \right)_{\mathrm{EM}}=\frac{\pi a_{em}^2 {\mu_{eff}}^{2}\,Z^{2}}{m_{e}^{2}}\left(\frac{1-T/E_{\nu}}{T} + \frac{T}{4 E_\nu^2} \right) F_{Z}^{2}(q^{2})\,.
\label{NMM-cross section}
\end{equation}
Note, that additional corrections are incorporated in Eqs.~(\ref{SM_dT}),~(\ref{NMM-cross section}), compared to Ref.~\cite{Kosmas:2015sqa}.

If the neutrino is of Dirac-type as in the SM, then the magnetic
moment is undetectably small due to its proportionality to the
neutrino mass. Even though from oscillation data the latter is
well-known to be small, one cannot rule out the possibility of
sizeable neutrino magnetic moments.
Indeed, in general scenarios where neutrinos are Majorana fermions, as
expected on general grounds, larger transition magnetic moments are
possible. For example, relatively sizeable contributions may be
predicted in models involving NSI~\cite{Papoulias:2015iga}. From
experimental perspectives, a potential signal will be detected as a
distortion of the nuclear recoil spectrum at very low energies where
the EM cross section dominates due to its $\sim 1/T$ dependence.  For
this reason, such challenging technological constraints require
innovative experimental advances towards reducing the threshold to the
sub-keV region.

Despite having vanishing electric charge, the first
  derivative in the expansion of the neutrino electric form factor
  entering the decomposition of the leptonic matrix element may
  provide non-trivial information concerning other neutrino electric
  properties~\cite{Broggini:2012df}. More specifically, the
  gauge-invariant definition of the neutrino root mean square
  charge-radius $\langle r^2_{\nu_{\alpha}}\rangle$, $\alpha=\{e,\mu,
  \tau\}$~\cite{Bardeen:1972vi,Lee:1973fw} leads to corrections to the
  weak mixing
  angle~\cite{Barranco:2007ea,Hirsch:2002uv,bernabeu:2000hf}
\begin{equation}
\sin^2 \theta_W \rightarrow \sin^2 \overline{\theta_W} + \frac{\sqrt{2} \pi a_{em}}{3 G_F} \langle r^2_{\nu_{\alpha}}\rangle\, .
\label{rve:definition}
\end{equation}

\section{The TEXONO experiment}    

In the present work we explore how well one can probe neutrino EM
phenomena with the TEXONO
experiment~\cite{Wong:2006nx,Deniz:2009mu,Soma:2014zgm} through
low-energy CENNS measurements near the Kuo-Sheng Nuclear Power
Station. Towards this purpose, the TEXONO Collaboration has pursued a
research program aiming at detecting neutrino-nucleus events by using
high purity Germanium-based detectors HPGe with sub-keV
threshold~\cite{Wong:2005vg,Wong:2010zzc,Chen:2014ypv}. 
%
%
According to the proposal, we consider a 1~kg $^{76}$Ge-detector operating with a threshold as low as $T_{thres}=100~\mathrm{eV_{ee}}$.
Due to the absence of precise
information regarding the fuel composition of the reactor core, we
only include the dominant $^{235}$U component of the anti-neutrino
spectrum. In this respect, for the present study we assume a typical
neutrino flux of $\Phi_{\bar{\nu}_e} = 10^{13}$ $\nu \, \mathrm{
  s^{-1} \, cm^{-2}}$ for a detector location at 28~m from the reactor
core.
In order to estimate the reactor antineutrino energy-distribution
$\eta_{\bar{\nu}_e}(E_\nu)$ for energies above 2~MeV, existing
experimental data from Ref.~\cite{Mueller:2011nm} are employed. We
stress that the main part of reactor anti-neutrinos is released with
energies $E_{\bar{\nu}_{e}}<2$~MeV, thus their contribution is crucial
and must be taken into account.  For their description we adopt the
theoretical estimates given in Ref.~\cite{Kopeikin:1997ve}.  
This will bring about improved sensitivities on the neutrino magnetic moment. 
%
\begin{figure}[h]
\begin{minipage}[b]{0.48\textwidth}
\centering
\includegraphics[width=\textwidth]{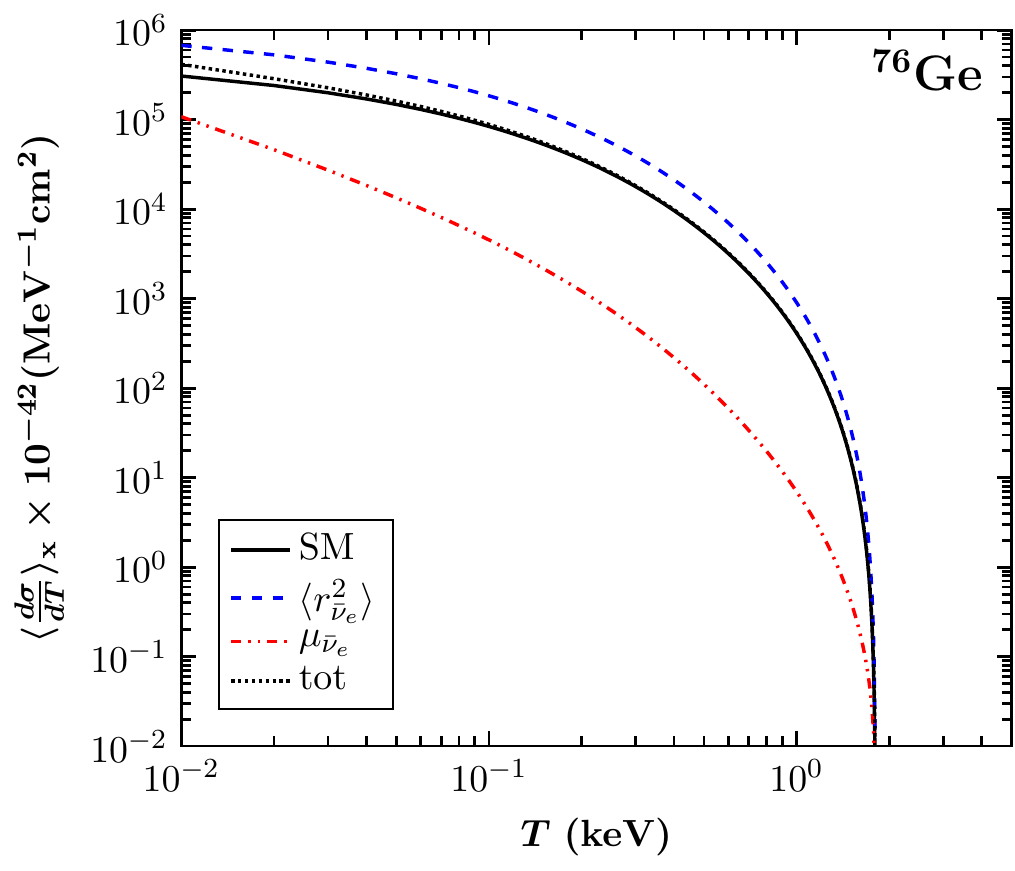} 
\end{minipage}
\hfil
\begin{minipage}[b]{0.48\textwidth}
\centering
\includegraphics[width=1.02\textwidth]{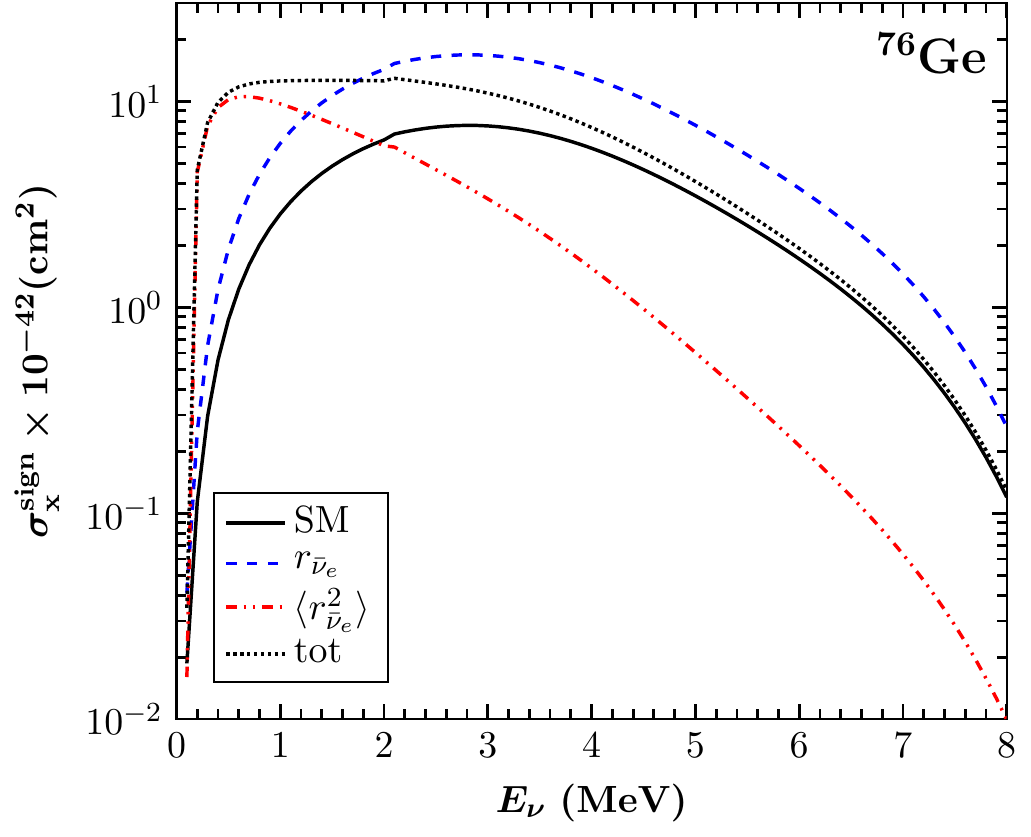}
\end{minipage}
\caption{Weak and electromagnetic differential (\textit{left panel})
  and total (\textit{right panel}) cross sections convoluted with
  reactor $\bar{\nu}_e$-spectra.}
\label{fig.0}
\end{figure}
%

\section{Results and discussion}

\subsection{Signal cross sections}

First we present and discuss the individual weak and electromagnetic
differential and total CENNS cross sections (see
Eq.~\ref{tot-crossec}) weighted over experimental reactor
anti-neutrino spectra~\cite{Mueller:2011nm,Kopeikin:1997ve}.  These
convoluted cross sections determine the neutrino signals expected to
be recorded at a nuclear detector (e.g. the $^{76}$Ge of the TEXONO
experiment).  For each interaction channel $x$,
$[x=\mathrm{SM,EM,tot}]$ the energy-integrated differential cross
section, $\langle d \sigma / dT \rangle_x$, is defined as
\begin{equation}
\langle \frac{d \sigma}{dT} \rangle_x = \int d E_{\nu}\,  \left( \frac{d \sigma}{dT}(E_\nu,T)\right)_x \eta_{\bar{\nu}_e}(E_\nu) \, ,
\end{equation}
where $\eta_{\bar{\nu}_e}(E_\nu)$ denotes the normalised neutrino energy-distribution. The corresponding signal cross section $\sigma^{\mathrm{sign}}_x$ reads
\begin{equation}
\sigma^{\mathrm{sign}}_x (E_\nu)= \int dT \,  \left( \frac{d \sigma}{dT}(E_\nu,T)\right)_x \eta_{\bar{\nu}_e}(E_\nu) \, .
\end{equation}
For the $^{76}$Ge detector, assuming the experimental constraints
placed recently by TEXONO ($\mu_{\bar{\nu}_e} = 7.4 \times 10^{-11}\,
\mu_B$~\cite{Wong:2006nx} and $\langle r^2_{\bar{\nu}_e} \rangle = 6.6
\times 10^{-32} \mathrm{cm^2}$~\cite{Deniz:2009mu}), the computed
results are illustrated in Fig.~\ref{fig.0}. One sees that in the case of
$\sigma^{\mathrm{sign}}$, the curve involving neutrino magnetic moment
contribution exceeds that of the pure SM weak rate at low neutrino
energies, $E_\nu$.
The curve containing neutrino charge radius contributions through
Eq.~\ref{rve:definition} is showing a similar behaviour as the pure SM
(the photon propagator cancellation leads to 4-fermion contact interaction~\cite{Chen:2014ypv,Giunti:2015gga}).

\subsection{Statistical analysis}

Our present analysis is strongly based on the estimation of the number
of CENNS events. Therefore, we first provide a brief description of
the conventions and approximations we use in our calculations. For
each interaction channel $x$, the number of CENNS events above a 
minimum nuclear recoil-energy, $T_{min}$, reads~\cite{Barranco:2005yy}
\begin{equation}
N_{x}= K \int_{E_{\nu_\mathrm{min}}}^{E_{\nu_\mathrm{max}}} \eta_{\bar{\nu}_e}(E_{\nu})\,dE_{\nu}\int_{T_{\mathrm{min}}}^{T_{\mathrm{max}}} \left( \frac{d \sigma}{dT}(E_{\nu},T) \right)_{x}\, d T\, .
\end{equation}
%
%
\begin{figure}[t]
\centering
\includegraphics[width=0.6\linewidth]{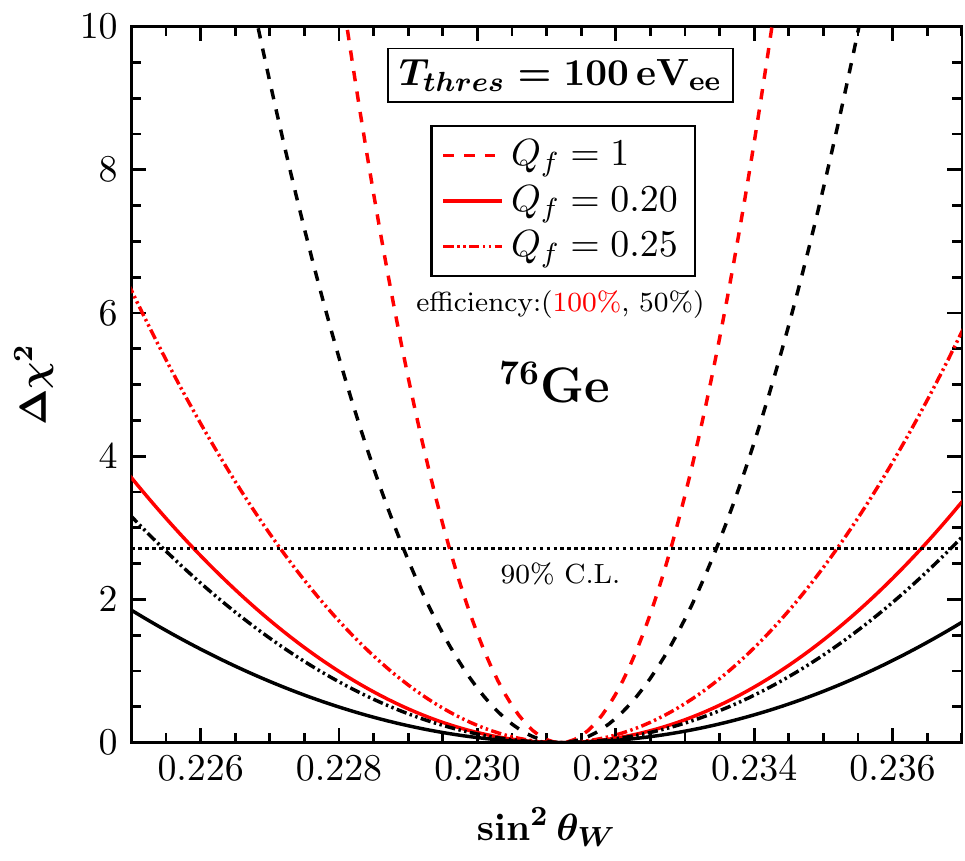}
    \caption{$\Delta \chi^2$ sensitivity profile as a function of the 
  weak mixing angle $\theta_W$ at TEXONO. The results are presented 
  for two detector efficiencies 100\% (50\%) shown with red (black)  
  colour and three values of the quenching factors $Q_f=(1, 0.20, 0.25)$ indicated with (dashed, solid, dashdotted) lines. (For interpretation of the references to colour in this figure legend, the reader is referred to the web version of this article.)
   }
\label{fig.1}
\end{figure}
%
%
\begin{table*}[ht!]
\centering
\caption{Summary of the sensitivities obtained for $\sin^2 \theta_W$
  (1$\sigma$) and for the EM neutrino parameters (90\% C.L.) at the
  TEXONO experiment. The results refer to various sensitivities 
  and 
  quenching factors. Comparing with Ref.~\cite{Kosmas:2015sqa} one
  sees that a substantial improvement in the sensitivity for the weak
  mixing angle $\sin^2 \theta_W$, the magnetic moment
  $\mu_{\bar{\nu}_e}$ parameter and the neutrino charge-radius
  $\langle r_{\bar{\nu}_e}^2 \rangle$ w.r.t. the COHERENT proposal.}
\resizebox{\textwidth}{!}{
\begin{tabular}{lccccccc}
\hline \hline

 & COHERENT~\cite{Kosmas:2015sqa}  &  \multicolumn{6}{c}{{TEXONO (this work)}}\\

(Target, Threshold) & (100~kg $^{76}\mathrm{Ge}$, $10~\mathrm{keV_{ee}}$) & \multicolumn{6}{c}{(1~kg $^{76}\mathrm{Ge}$, $100~\mathrm{eV_{ee}})$}\\
 
 \cmidrule(lr{4pt}){2-2} \cmidrule(lr){3-8}

Efficiency & 67\% & \multicolumn{3}{c}{100\%}   & \multicolumn{3}{c}{50\%} \\

\cmidrule(lr{4pt}){3-5} \cmidrule(lr){6-8}

Quenching  & $Q_f=1$ & $Q_f=1$ & $Q_f=0.20$ & $Q_f=0.25$ & $Q_f=1$ & $Q_f=0.20$ & $Q_f=0.25$ \\
\hline

$\delta s^2_W(\bar{\nu}_{e})$         &  0.0055 &  0.0010 & 0.0033 & 0.0025  & 0.0014 & 0.0046 & 0.0035\\
Uncer. (100\%) & 2.36 & 0.43 & 1.41 & 1.08 & 0.61 & 1.97 & 1.51 \\

 $\mu_{\bar{\nu}_e} \times 10^{-10}\, \mu_B$ & 9.46 & 0.40 & 0.98 & 0.83 & 0.47 & 1.17 & 0.99 \\

$\langle r_{\bar{\nu}_e}^2 \rangle \times 10^{-32}\, \mathrm{cm^2}$   & -0.38 -- 0.37 & -0.07 -- 0.07 & -0.22 -- 0.22 & -0.17 -- 0.17 & -0.10 -- 0.10 & -0.32 -- 0.31 & -0.24 -- 0.24\\
 \hline \hline
\end{tabular}
}
\label{table.summary}
\end{table*}
%
In the above expression, $K = N_{\mathrm{targ}} t_{\mathrm {tot}}
\Phi_{\bar{\nu}_e}$, with $N_{\mathrm{ targ}}$ being the total number
of atoms in the detector and $t_{\mathrm{ tot}}$ the relevant
irradiation period. Note that potential effects due to neutrino
oscillation in propagation are neglected, since this is well satisfied
for the short-baselines considered here. 
%
%
The numerical results throughout this work refer to a 1~kg $^{76}$Ge-detector, one year of data taking and a detector threshold of $100~\mathrm{eV_{ee}}$. In addition, we consider two different detector efficiencies including an optimistic approach of a perfectly efficient detection capability and the more realistic scenario assuming a recoil acceptance of 50\%.

The present calculations take into consideration the fact that the nuclear recoil events are quenched~\cite{Simon:2002cw} (in Ref.~\cite{Kosmas:2015sqa}, where for each target the calculation is referred to the
nuclear recoil-energy window~\cite{Bolozdynya:2012xv}, such a treatment is not necessary). Corrections of this type are crucial since for a given ionisation detector the observed energy (equivalent to an electron energy) is lower than the total nuclear recoil-energy, i.e. much energy is converted to heat (phonons) which is not measured, especially at low energies~\cite{Giomataris:2005fx}. To convert from nuclear recoil-energy ($\mathrm{eV_{nr}}$) to electron equivalent energy ($\mathrm{eV_{ee}}$), we multiply the energy scale by a quenching factor, $Q_f$. In principle $Q_f$ varies with the nuclear-recoil energy~\cite{Vergados:2009ei} and has to be determined experimentally, however for the sub-keV Germanium-based targets considered here it can be well-approximated as constant with typical values in the range 0.20-0.25~\cite{Wong:2005vg}. Thus, the TEXONO threshold $T_{thres}= 100~\mathrm{eV_{ee}}$ corresponds to nuclear recoil-energy $T_{\mathrm{min}}= 500~\mathrm{eV_{nr}}$ for $Q_f=0.20$ and $T_{\mathrm{min}}= 400~\mathrm{eV_{nr}}$ for $Q_f=0.25$, correspondingly the maximum nuclear recoil-energy $T_{\mathrm{max}}=1.81~\mathrm{keV_{nr}}$, is restricted to a maximum observable energy of 362 and $\mathrm{452~eV_{ee}}$. 

%
\begin{figure}[t]
\begin{minipage}[b]{0.48 \textwidth}
\centering
\includegraphics[width=\textwidth]{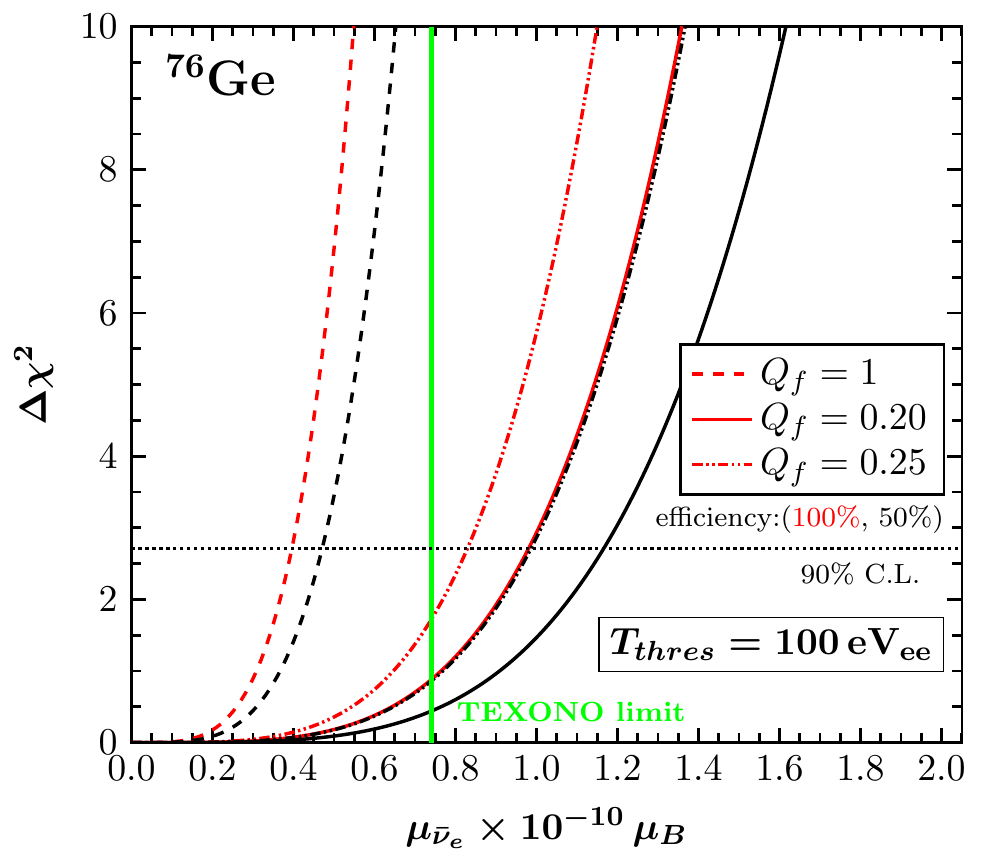}
\end{minipage}
\hfill
\begin{minipage}[b]{0.48 \textwidth}
\centering
\includegraphics[width=1.05\textwidth]{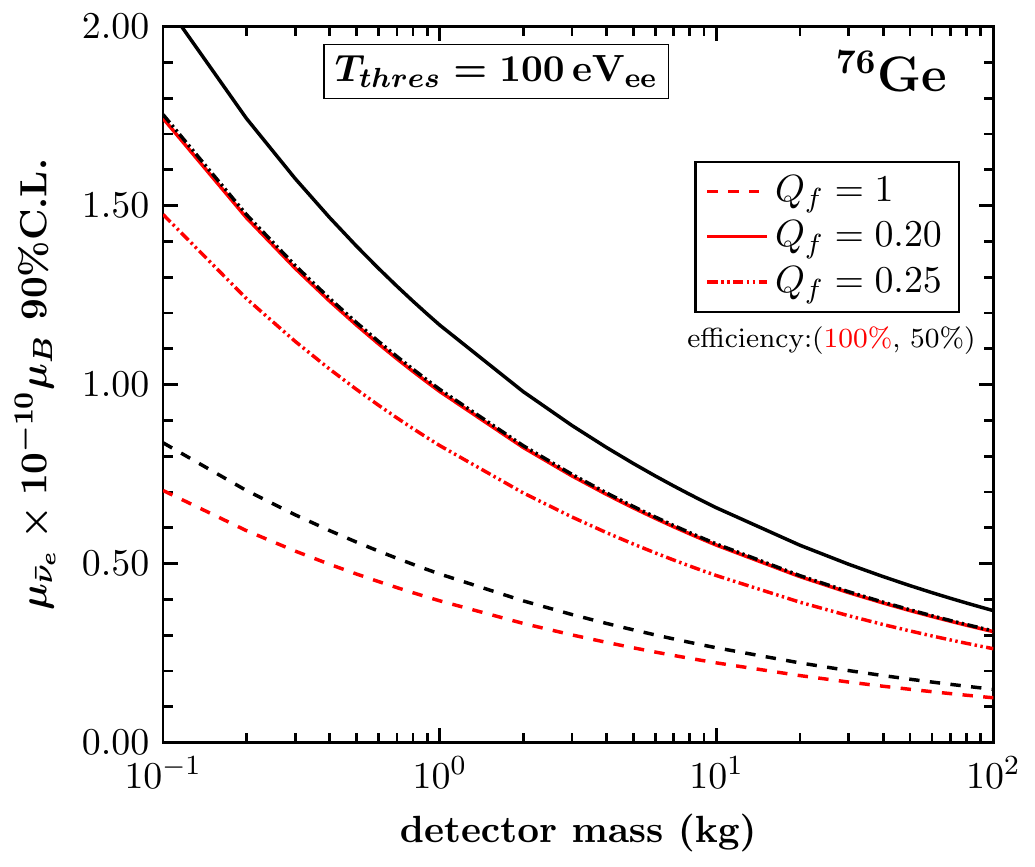}
\end{minipage}
\caption{
 (\textit{Left panel}) $\Delta \chi^2$ profiles in terms of the
neutrino magnetic moment parameter $\mu_{\bar{\nu}_{e}}$ in units of
$10^{-10} \mu_B$ at TEXONO. (\textit{Right panel}) Sensitivity to $\mu_{\bar{\nu}_{e}}$ 
at 90\% C.L. as a function of the detector mass. Same conventions 
as in Fig.~\ref{fig.1} are used.
 }
\label{fig.2}
\end{figure}
%
%
\begin{figure*}[t]
\centering
\includegraphics[width= 0.90 \textwidth]{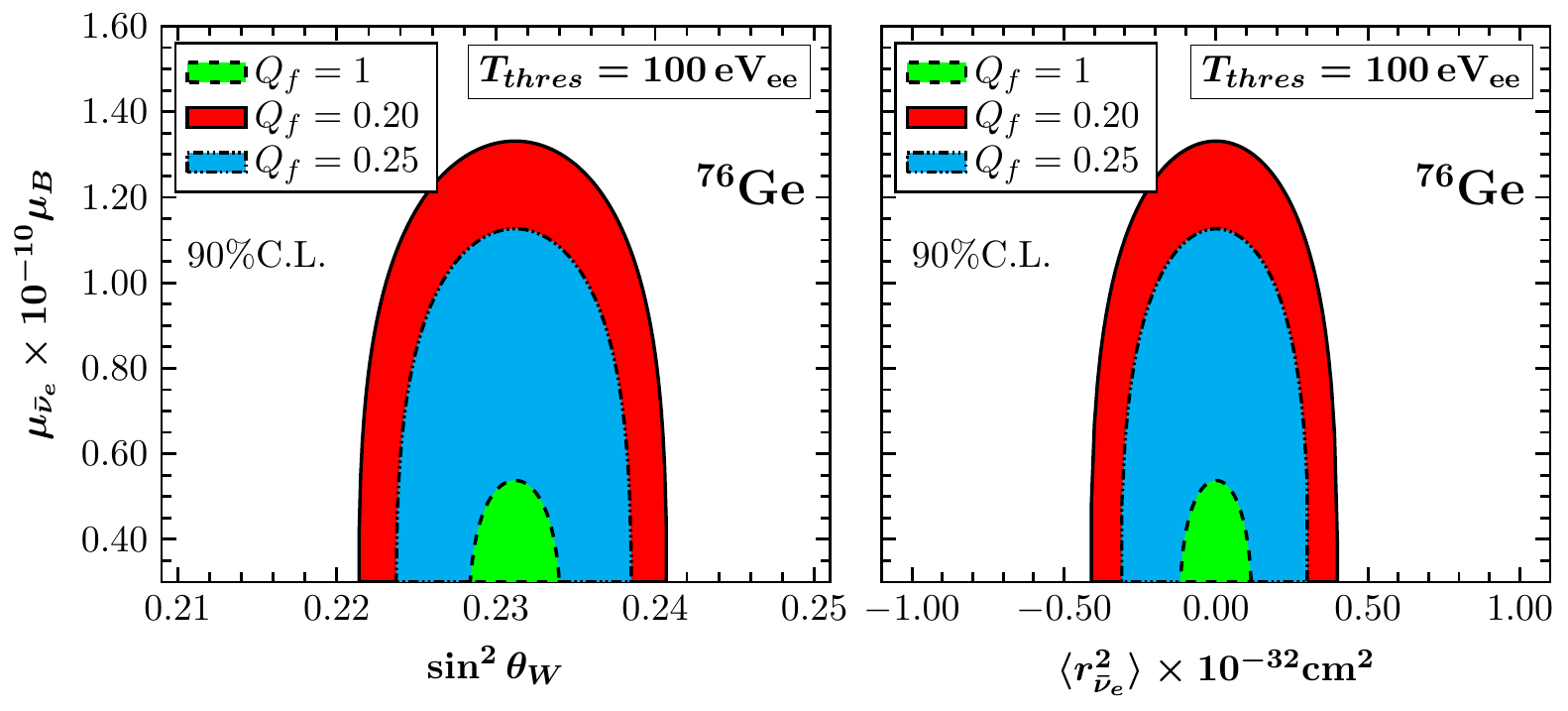}
\caption{90\% C.L. allowed regions in the ($\sin^2
  \theta_W$-$\mu_{\bar{\nu}_{e}}$) plane (\textit{left panel}) and the
  ($\langle r_{\bar{\nu}_e}^2 \rangle$-$\mu_{\bar{\nu}_{e}}$) plane
  (\textit{right panel}) from a two parameter combined analysis. See the text
  for more details.}
\label{fig.3}
\end{figure*}
%

The sensitivity of the TEXONO experiment to the SM weak mixing angle
$\sin^2 \theta_W$ is quantitatively determined on the basis of a
$\chi^2$-type analysis relying on statistical errors only~\cite{Kosmas:2015sqa}
\begin{equation}
  \chi^2 = \left( \frac{N_{SM} - N_{SM}(\sin^2 \theta_W)}{\delta N_{SM}} \right)^2\, .
\label{chi.sw}  
\end{equation}
In our calculational procedure we have assumed that the TEXONO
experiment will detect the precise number of SM events, $N_{\rm{SM}}$,
by fixing the electroweak mixing parameter to the PDG value
i.e. $\hat{s}^2_Z=0.23120$~\cite{Beringer:1900zz}.
%
%
We estimate the expected events $N_{\mathrm{SM}}=(27962, \,2586, \,4415)$  assuming  $Q_f=(1, \,0.20, \,0.25)$ and a detection threshold $\mathrm{100~eV_{ee}}$, in good agreement with Ref.~\cite{Wong:2010zzc}. 
The $\chi^2$ function is then minimised with
respect to $\sin^2 \theta_W$, by varying this parameter around its
central value, taken as the value reported by the PDG.

We have explicitly verified that there are good prospects for making
precision tests of the SM by using low-energy $^{76}$Ge detectors.
Our results for the TEXONO sensitivity to the weak mixing angle are
presented in Fig.~\ref{fig.1}.  Furthermore, we have also evaluated
the 1$\sigma$ error band on $\sin^2\theta_W$ defined as $\delta \sin^2
\theta_W \equiv \delta s^2_W=(s^2_{W^{max}}-s^2_{W^{min}})/2$ as well
as the corresponding uncertainty $\delta s^2_W/ \hat{s}^2_Z$, with
$s^2_{W^{max}}$ ($s^2_{W^{min}}$) being the respective upper (lower)
bound.  The resulting sensitivities are shown in
Table~\ref{table.summary}. Specifically, neglecting the quenching corrections ($Q_f=1$), the improvement upon previous
results~\cite{Kosmas:2015sqa} is up to 82\% (74\% when realistic
efficiencies are taken into account). Furthermore, the effect for $Q_f=0.20~(0.25)$ leads to reduction of $\delta \sin^2 \theta_W$ sensitivity by a factor of 3.3~(2.5) for both detection efficiencies.

Prompted by the upcoming generation of low-threshold nuclear
detectors, we have made an effort to identify possible deviations from
the SM neutrino-quark interaction cross section originated by
non-standard neutrino EM properties.  In particular, analysing their
sensitivity to electromagnetic CENNS events we have found that
important deviations may be induced by the presence of a non-zero
transition neutrino magnetic moment $\mu_{eff} \equiv
\mu_{\bar{\nu}_e}$.
  In order to determine this sensitivity we use a $\chi^2$ function of
  the form~\cite{Tortola:2004vh}
\begin{equation}
  \chi^2 = \left( \frac{N_{\rm{SM}} - N_{\text{tot}}(\mu_{\bar{\nu}_{e}})}{\delta N_{\rm{SM}}} \right)^2\, .
  \label{NMM-chi}
\end{equation}

In Eq.~(\ref{NMM-chi}), we substitute the SM cross section by the one
given in Eq.~(\ref{tot-crossec}) in order to account for possible
events, $N_{\text{tot}}$, originating from the corrections associated
to the non-trivial structure of the neutrino EM current, as discussed
previously. The corresponding results obtained by varying the
effective transition neutrino magnetic moment, $\mu_{\bar{\nu}_e}$,
are presented in Fig.~\ref{fig.2} (left panel). The experimental
TEXONO limit from $\bar{\nu}_e-e$ scattering, $\mu_{\bar{\nu}_e} = 7.4
\times 10^{-11} \mu_B$~\cite{Wong:2006nx}, is also shown for
comparison (the most stringent bound on the neutrino magnetic
moment comes from the reactor experiment GEMMA as $\mu_{\bar{\nu}_e} =
2.9 \times 10^{-11} \mu_B$ at 90\% C.L.~\cite{Beda:2013mta}).  One
sees that the prospects are very promising.  Indeed, from
Table~\ref{table.summary}, we find that the attainable sensitivities
are improved by about one order of magnitude compared to the
corresponding expectations at a SNS facility, considered recently
in~\cite{Kosmas:2015sqa}. Note however, that experiments at the SNS
are not optimised to measure electron-neutrino properties.
    
Moreover, it is worth mentioning that for the case of a $\mathrm{100~eV_{ee}}$  threshold and $Q_f=1$, the resulting sensitivity is by 46\% (36\% for the case of realistic efficiency) better than the existing limits derived from $\bar{\nu}_e-e$ scattering TEXONO data~\cite{Wong:2006nx}.
However, assuming $Q_f=0.20~(0.25)$ the above sensitivity reduces by a factor of 2.5~(2.1) for both recoil acceptances.
 From our calculations we have also found that neglecting quenching corrections the sensitivity to $\mu_{\bar{\nu}_e}$ of a given detector with mass $m$ is roughly equivalent to that of a detector with ten times bigger mass for the case of  $Q_f=0.25$. The results concerning this point are shown in Fig.~\ref{fig.2}
(right panel).

In view of our previous discussion, the TEXONO sensitivity to
$\langle r^2_{\bar{\nu}_e}\rangle$-related searches is estimated
through the definition of the $\chi^2$ given in Eq.~(\ref{chi.sw}) by
replacing $\sin^2 \theta_W$ with that of Eq.~(\ref{rve:definition}) and
fixing $\sin^2 \overline{\theta_W}$ to the PDG value.
After the $\chi^2$ minimisation we find that the TEXONO experiment is
expected to be very sensitive to EM contributions of this type. The
estimated 90\% C.L. sensitivities are presented in Table~\ref{table.summary}.  In the particular case of $Q_f=0.25$ we see that thanks to the observation of CENNS events, TEXONO can reach an improvement of the order of 35\% or more, with respect to similar calculations~\cite{Kosmas:2015sqa}. Again, this sensitivity reduces by a factor of 3.2~(2.4) when quenching corrections $Q_f=0.20~(0.25)$ are taken into account. Moreover, it is worth noting that the latter sensitivities are by one order of magnitude better than the
current TEXONO constraint obtained from $\bar{\nu}_e-e$
scattering~\cite{Deniz:2009mu}.

Finally, it is also interesting to show the combined sensitivities
obtained by varying two of the above parameters ($\sin^2 \theta_W$,
$\mu_{\bar{\nu}_{e}}$ and $\langle r_{\bar{\nu}_e}^2 \rangle$)
simultaneously.  The 90\% C.L. allowed regions in the ($\sin^2
\theta_W$-$\mu_{\bar{\nu}_{e}}$) and ($\langle r_{\bar{\nu}_e}^2
\rangle$-$\mu_{\bar{\nu}_{e}}$) plane are shown in the left and right
panel of Fig.~\ref{fig.3}, respectively for different quenching factors.  One notices that the
resulting parameter space is substantially reduced with respect to the
corresponding sensitivity regions for muon neutrinos at a SNS experiment (see e.g. Ref.~\cite{Kosmas:2015sqa}). The
latter is a direct consequence of the low-threshold TEXONO detectors
adopted in the present study.
  
\section{Summary and conclusions}

In this work we have explored the possibility of performing Standard
Model precision studies and probing for new physics through low energy
neutral-current coherent elastic neutrino-nucleus scattering (CENNS)
at the TEXONO experiment. Moreover, we have presented a comprehensive
analysis for the case of potential sub-leading neutrino EM
interactions. The calculated convoluted cross sections, clearly indicate the need for novel detector technologies with sub-keV sensitivities. Furthermore, from a nuclear physics point of view, the reactor neutrino beam induces transitions in the bound nuclear spectrum while the Spallation Neutron Source (SNS) beam may excite in addition much higher transitions of the nuclear detector.
We conclude that, apart from providing the first ever detection of
CENNS events, low threshold Germanium-based kg-scale detectors, e.g.
TEXONO, will bring substantial improvements on precision SM tests as
well as sensitivities on neutrino EM properties, such as the
neutrino magnetic moment and the neutrino charge radius.
We show explicitly that the sensitivities improve by up to one order
of magnitude with respect to previous estimates.
In this paper we have used realistic nuclear structure calculations
within the context of the quasi-particle random phase approximation
(QRPA) and a simple $\chi^2$-type analysis taking into account quenching effects.
We have also checked that our sensitivities are determined mainly by
the number of events: a binned sample would result in differences
less than percent.
It is also worth mentioning that a global fit including an experiment
at the SNS added to the TEXONO experiment
will not essentially improve these results, since SNS provides the
best sensitivities for the case of muonic neutrinos.
Hence, the two experiments are clearly complementary.

\section*{Acknowledgements}
\sloppy
This work was supported by Spanish MINECO under grants FPA2014-58183-P,
Multidark CSD2009-00064 (Consolider-Ingenio 2010 Programme), by EPLANET, by the CONACYT grant 166639
(Mexico) and the PROMETEOII/2014/084 grant from Generalitat
Valenciana.  DKP was supported by \textit{"Prometeu per a grups d'
  investigaci\'o d' Excel$\cdot$l\`{e}ncia de la Conseller\'{i}a d'
  Educaci\'o, Cultura i Esport, CPI-14-289" (GVPROMETEOII2014-084)}.
MT is also supported by a Ramon y Cajal contract of the Spanish MINECO (grant SEV-2014-0398
(MINECO)). TSK wishes to thank Prof. H.T. Wong for the useful discussion on TEXONO experiment during attending the NDM-15 conference. DKP is indebted to Prof. J.D. Vergados for stimulating discussions.




\end{document}